\documentclass[twocolumn,aps,showpacs,prb]{revtex4-1}

\usepackage{amsmath,amssymb,graphics,epsfig,epstopdf,color,multirow,array,verbatim,ulem,braket,tabularx}
\usepackage[colorlinks,linkcolor=blue,citecolor=blue,urlcolor=blue]{hyperref}
\usepackage{color}
\usepackage{graphicx}
\usepackage{epstopdf}
\usepackage{amsmath}
\usepackage{multirow}
\usepackage{ulem}

\begin{document}

\title{Large anomalous Hall effect induced by gapped nodal lines in GdZn and GdCd}

\author{Ning-Ning Zhao}
\author{Kai Liu}\email{kliu@ruc.edu.cn}
\author{Zhong-Yi Lu}\email{zlu@ruc.edu.cn}

\affiliation{Department of Physics and Beijing Key Laboratory of Opto-electronic Functional Materials $\&$ Micro-nano Devices, Renmin University of China, Beijing 100872, China}

\date{\today}

\begin{abstract}

The topological properties and intrinsic anomalous Hall effect of CsCl-type ferromagnets GdZn and GdCd have been studied based on first-principles electronic structure calculations. According to the calculated band structures, both GdZn and GdCd host nodal lines near the Fermi level. Once the magnetization breaks the mirror symmetries, the nodal lines are gapped. This can create a huge Berry curvature. A large anomalous Hall effect is then generated when the Fermi level resides within the opened gaps of the nodal lines. Our works indicate that GdZn and GdCd can provide a promising platform for exploring the interplay between topological property and magnetism.

\end{abstract}


\maketitle

\section{INTRODUCTION}

The anomalous Hall effect (AHE) is a fascinating electronic transport property \cite{AHE}. It is characterized by the resistivity acquired via the magnetization of materials in the direction perpendicular to the external electric field \cite{hall-1,book}. So far, three mechanisms have been proposed for the AHE.  Two of them claim that the AHE results from the extrinsic contributions that include the skew scattering and the side jump \cite{hall-2,jump}: The former is considered as the horizontal charge accumulation due to the asymmetric scattering by the impurity, while the latter is described as a lateral jump of the electronic coordinates after scattering. Another mechanism points out that the AHE is an intrinsic effect that is closely related to the Berry curvature of the electronic Bloch states \cite{Luttinger,A. H. MacDonald,QAHE,Fe,BC}, as exemplified in many ferromagnetic or ferrimagnetic materials. In this case, the magnetization breaks the time-reversal symmetry, then results in a nonzero Berry curvature, and in turn generates an intrinsic anomalous Hall conductance (AHC) \cite{BP,advance,Co2VGa}.

The coexistence of magnetism and topological band structure in a material may lead to a strong AHE. If a ferromagnetic material possesses a nontrivial
topological band structure near the Fermi level, such as Weyl nodes or gapped nodal lines, there will be usually a large Berry curvature and a huge AHC \cite{nodes,weyl,Fe2MnP}. A Weyl node can be considered as a magnetic monopole while a nodal line can be regarded as a magnetic vortex line in  momentum space \cite{magnetic monopoles,Fe3GeTe2}. The nodal lines protected by crystal symmetry in the absence of spin-orbital coupling (SOC) may evolve into a set of Weyl nodes with the inclusion of SOC. Recently, a large intrinsic AHE has been observed in ferromagnetic Weyl semimetals and nodal line semeimetals,  originating from the nontrivial band topologies \cite{Co3Sn2S2-1,Co3Sn2S2-2,Fe3GeTe2,Co2MnAl}. Therefore, searching for a ferromagnetic material with nontrivial topological band structure near the Fermi level is of great significance for realizing the large AHE\cite{spintronics}.

Based on first-principles electronic structure calculations, here we predict that the CsCl-type Gd-based intermetallics GdZn and GdCd are a new class of ferromagnets with the large anomalous Hall effect due to the gapped nodal lines. Previous experiments have shown that both GdZn and
GdCd are isotropic ferromagnets \cite{isotropic-GdZn,isotropic-GdCd} with high Curie temperatures of 270 K \cite{GdZn-TC} and 265 K\cite{GdCd-TC}, respectively. Our calculations thus uncover the promising materials with the AHE near the room temperature.

\section{Method}

We have investigated the electronic structures and transport properties of GdZn and GdCd based on the first-principles electronic structure calculations\cite{DFT}. The projected augmented wave method \cite{PAW} as implemented in the Vienna \textit{ab initio} simulation package (VASP)\cite{vasp1,vasp2} was employed to perform the calculations. The generalized gradient approximation (GGA)\cite{GGA} of Perdew-Burke-Ernzerhof (PBE) type was used for the exchange-correlation functional. The kinetic energy cutoff of 360 eV was utilized for the plane wave basis. A $15 \times 15 \times 15$ \textbf{k}-point mesh was used for the Brillouin zone (BZ) sampling and the Gaussian smearing method with a width of 0.05 eV was adopted for the Fermi surface broadening. The generalized gradient approximation plus Hubbard U (GGA+U) approach \cite{GGA+U} with U$_\text{eff}=6.0$ eV\cite{Gd-U} was carried out to treat the electronic correlation effect among Gd 4$f$ electrons. Both cell parameters and internal atomic positions were fully relaxed until the forces on all atoms were smaller than 0.01 eV/{\AA}. The SOC effect was taken into account in the  calculations of electronic structures. To explore the nontrivial band topology and the intrinsic AHE, the tight-binding Hamiltonian was
constructed with the maximally localized Wannier functions\cite{WF,w90} for the outmost $s$, $p$, $d$ and $f$ orbitals of Gd atoms and the outmost $s$, $p$, and $d$ orbitals of Zn (Cd) atoms generated by the first-principles calculations. Based on the tight binding Hamiltonian, the AHC and the Berry curvature were evaluated via the Kubo-formula approach in the linear response scheme\cite{BP}as follows,
\begin{equation}\label{}
\sigma_{xy} = -\frac{e^{2}}{\hbar}\int_{BZ}\frac{d^{3}\textbf{k}}{(2\pi)^{3}}\sum_{n}f_{n}(\textbf{k})\Omega_{n}^{z}(\textbf{k}),
\end{equation}

\begin{equation}\label{}
\Omega_{n}^{z}(\textbf{k})=-2\rm{Im}\sum_{m\neq n }\frac{\bra{\psi_{n\textbf{k}}} v_{x} \ket{\psi_{m\textbf{k}}} \bra{\psi_{m\textbf{k}}} v_{y} \ket{\psi_{n\textbf{k}}}}{[E_{m}(\textbf{k})-E_{n}(\textbf{k})]^{2}},
\end{equation}
where $f_{n}(\textbf{k})$ is the Fermi-Dirac distribution function, $n$ is the index of the occupied bands, $E_{n}(\textbf{k})$ is the eigenvalue of the $n$th eigenstate $\psi_{n}(\textbf{k})$, and $v_{i}=\frac{1}{\hbar}\frac{\partial H(\textbf{k})}{\partial \textbf{k}_{i}} $ is the velocity operator along the $i$ ($i=x, y, z$) direction. A \textbf{k}-point grid of $301\times301\times301$ was used in the integral calculation. The surface states in the projected two-dimensional (2D) BZ were obtained from the surface Green's function of the semi-infinite system with the WANNIERTOOLS package\cite{WT}. To simulate the charge doping effect, the total electron number of the system was changed in the self-consistent and band-structure calculations.

\begin{figure}[tb]
\includegraphics[angle=0,scale=0.155]{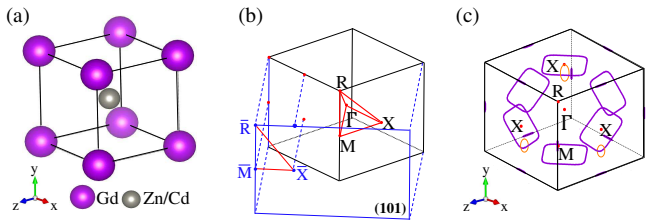}
\caption{(a) Crystal structure of the CsCl-type GdZn and GdCd. (b) Bulk Brillouin zone (BZ) and projected two-dimensional BZ of the (101) surface. The red lines indicate the high-symmetry paths in the BZ. (c) Nodal lines in the BZ without the spin-orbital coupling(SOC). The purple loops indicate the nodal rings in the $\bf{k}_{x,y,z}$=$\pm\pi$ planes and the orange circles indicate the closed paths wrapping the nodal lines for the Berry phase calculation.}
\label{fig1}
\end{figure}

\section{Result}

The gadolinium intermetallics GdZn and GdCd crystalize in the CsCl-type cubic structure with space group \textit{Pm}-3\textit{m} (No. 221). As shown in Fig. \ref{fig1}(a), Gd and Zn (Cd) atoms occupy the 1$\textit{a}$ (0.0 0.0 0.0) and 1$\textit{b}$ (0.5 0.5 0.5) Wyckoff positions, respectively. The optimized lattice constants are 3.606 {\AA} for GdZn and 3.774 {\AA} for GdCd respectively, which agree well with their measured values \cite{GdZn-lattice,GdCd-lattice}. Previous experiments indicate that they are both simple ferromagnets and have no anisotropy along three principle axes \cite{isotropic-GdZn,isotropic-GdCd}. We thus chose the Gd spins to be aligned along the [001] direction. The calculated local magnetic moments on Gd atoms are 7.276 $\mathrm{\mu_{B}}$ for GdZn and 7.287 $\mathrm{\mu_{B}}$ for GdCd, also in good accordance with the experimental results \cite{GdZn-TC,isotropic-GdCd}. Figure \ref{fig1}(b) displays the bulk BZ and the projected 2D BZ of the (101) surface along with the high-symmetry \textbf{k} points.

\begin{figure}[tb]
\includegraphics[angle=0,scale=0.25]{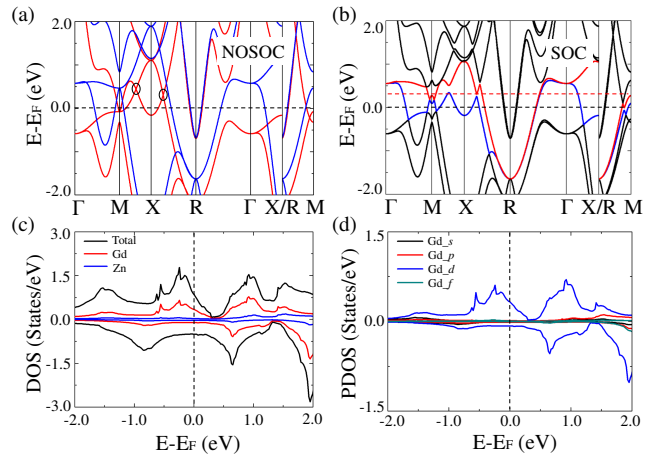}
  \caption{(a) Spin-resolved band structure of GdZn calculated without the spin-orbital coupling (SOC). The red and blue lines represent the spin-up and spin-down bands, respectively. (b) Band structure of GdZn calculated with the SOC. The magnetization sets along the [001] direction. The blue and red solid lines denote the highest valence and lowest conduction bands, respectively. (c) Density of states (DOS) and (d) Partial DOS of Gd atom for ferromagnetic GdZn calculated without SOC. The upper and lower parts are the spin-up and spin-down components, respectively. The Fermi level sets to zero.}
 \label{fig2}
\end{figure}

Figures \ref{fig2}(a) and \ref{fig2}(b) show the band structures of GdZn along the high-symmetry paths of the BZ calculated without and with the SOC, respectively. The magnetization direction of Gd spins is set parallel to the $z$ axis. There are several bands crossing the Fermi level, indicating that GdZn is a ferromagnetic metal. According to the calculated density of states [Figs. \ref{fig2}(c) and \ref{fig2}(d)], the 5$d$ orbitals of Gd atoms have major contributions near the Fermi level. In comparison, there is little contribution from the Gd $f$ orbitals in the energy interval of [-2, 2] eV with respect to the Fermi level.

A careful examination of the band structure of GdZn in Fig. \ref{fig2}(a) indicates that there is a band inversion in the spin-up channel around the $\mathrm{X}$ point at about 0.3 eV above the Fermi level. In the absence of SOC, the magnetization has no easy axis and the space group of GdZn is O$\mathrm{_{h}}$, which contains the mirror planes \textit{M$_{x}$}, \textit{M$_{y}$}, and\textit{ M$_{z}$}. With the protection of the mirror symmetries\textit{ M$_{i}$}($i=x,y,z$), the crossings of two bands form nodal rings in the corresponding $\textbf{k}$$_{i}=\pm\pi$ ($i=x,y,z$) planes [the purple loops in Fig. \ref{fig1}(c)]. The obtained Berry phase defined on a closed loop [the orange rings in Fig. \ref{fig1}(c)] wrapping  the nodal lines is $\pi$, which indicates the nodal line is topologically nontrivial\cite{TlTaSe2,nodal-line,YN}. Once the SOC is taken into account, the local spins couple with the crystal lattice and the system has an easy axis for the magnetization. As a result, the mirror plane perpendicular to the magnetization axis is preserved while the other mirror planes parallel to the magnetization axis are broken. With a z-axis magnetization, the band crossings between the highest valence band (blue solid line) and the lowest conduction band (red solid line) open a gap [Fig. \ref{fig2}(b)]around the X point in the $\textbf{k}$$_{x}=\pi$ plane[Fig.\ref{fig1}(b)].

\begin{figure}[tb]
\includegraphics[angle=0,scale=0.38]{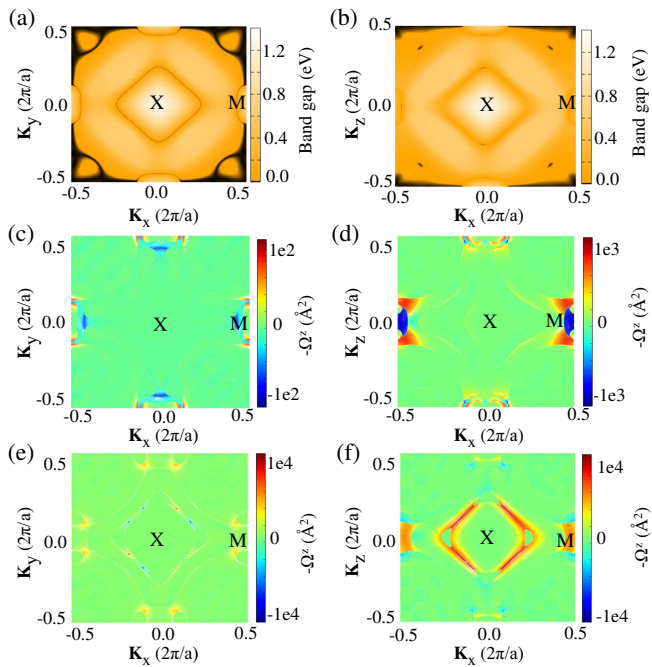}
  \caption{Band gaps of GdZn in the (a) $\textbf{k}$$_{z}=\pi$ plane and (b) $\textbf{k}$$_{y}=\pi$ plane with the magnetization along the [001] direction. Negative $z$ components of Berry curvatures -$\Omega^{z}(\textbf{k})$ for GdZn in the corresponding $\textbf{k}$$_{z}=\pi$ and $\textbf{k}$$_{y}=\pi$ planes with the magnetization along the [001] direction at fixed energies of (c)(d) E$_{f}$ and (e)(f) E$_{f}$+0.31 eV. }
  \label{fig3}
\end{figure}

When the magnetization is set along the [001] direction, the symmetry of the GdZn lattice is reduced to $C_{4h}$. Considering that the mirror operation \textit{M$_{\textbf{n}}$} can be obtained by the combination of the two-fold rotation $C_{2}^{\textbf{n}}$ and space inversion $I$: $M_{\textbf{n}}=C_{2}^{\textbf{n}}I$, the $C_{2x}$ and $C_{2y}$ symmetries in GdZn are broken by the magnetization along the [001] direction while the $C_{2z}$ symmetry is preserved. As a result, the mirror symmetry operations \textit{M$_{x}$} and \textit{M$_{y}$} are no longer allowed, while the mirror symmetry\textit{ M$_{z}$} remains. By checking the band gap in the $\textbf{k}$ plane as shown in Fig. \ref{fig3}(a), the nodal ring is visible as an oval black ring centered around the X point, which is protected by the preserved mirror\textit{ M$_{z}$} in the $\textbf{k}$$_{z}=\pi$ plane. On the contrary, the nodal ring is gapped out in the $\textbf{k}$$_{x}=\pi$ and $\textbf{k}$$_{y}=\pi$ planes due to the broken mirror symmetries of\textit{ M$_{x}$} and\textit{ M$_{y}$} [as examplified in Fig. \ref{fig3}(b)].

\begin{figure}[tb]
\includegraphics[angle=0,scale=0.35]{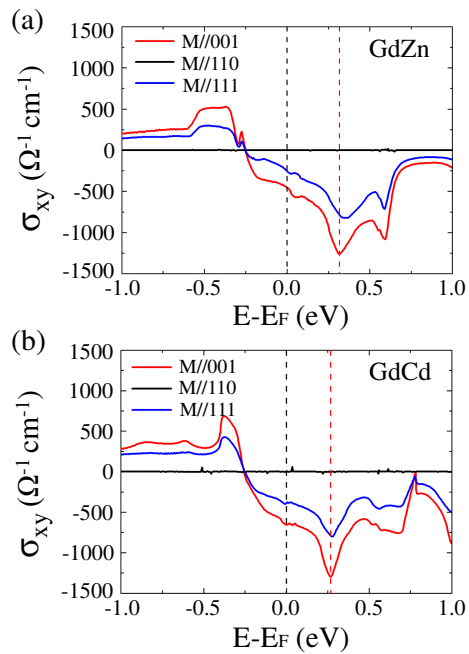}
  \caption {Calculated anomalous Hall conductance with different magnetization directions along the [001] (red solid line), [110] (black solid line), and [111] (blue solid line) directions for (a) GdZn and (b) GdCd, respectively.}
 \label{fig4}
\end{figure}

The nontrivial band structures usually generate huge Berry curvatures. Figures \ref{fig3}(c) and \ref{fig3}(d) show the negative $z$ components of Berry curvature distributions at the Fermi level in the $\textbf{k}$$_{z}$=$\pi$ and $\textbf{k}$$_{y}$=$\pi$ planes, respectively. There are hot peaks (about $10^{2}$) around the \textit{M} point, where the Fermi level lies in the band gap induced by the SOC. More importantly, when the Fermi level is located within the gap of the broken nodal line [Fig. \ref{fig2}(b)], the Berry curvature has a much larger magnitude (about $10^{3}$) around the X point in the $\textbf{k}$$_{y}$=$\pi$ plane as shown in Fig. \ref{fig3}(f). In contrast, the preserved gapless nodal ring in the $\textbf{k}$$_{z}$=$\pi$ plane contributes weak Berry curvature [Fig. \ref{fig3}(e)].

The strong Berry curvature may induce giant anomalous transport behavior. We have thus calculated the AHC by integrating the Berry curvature of the occupied bands in the whole BZ [Eqs. (1) and (2)]. As shown in Fig. \ref{fig4}(a), the calculated value of the intrinsic AHC reaches $-$453 $\Omega^{-1}$cm$^{-1}$ at the Fermi level, which is similar to that of the ferromagnetic nodal-line semimetal Fe$_{3}$GeTe$_{2}$\cite{Fe3GeTe2}. Remarkably, the gapped nodal rings lead to a large value of $-$1260 $\Omega^{-1}$cm$^{-1}$ at 0.315 eV above the Fermi level, which is comparable to that of the ferromagnetic Weyl semimetal Co$_{3}$Sn$_{2}$S$_{2}$\cite{Co3Sn2S2-1,Co3Sn2S2-2}. According to our estimation, a doping of $\sim$ 0.3 electrons per unit cell may shift the Fermi level into the gap of nodal line [red dashed line in Fig. \ref{fig2}(b)]. In comparison with the AHC of the [001] magnetization, we have also studied the AHC with the [110] and [111] magnetization axes. The AHC of the [111] magnetization is slightly lower than that of the [001] magnetization, while the value of the [110] magnetization is almost zero. The anisotropy of AHC originates from the different crystal symmetries which are determined by the magnetization axis \cite{EuB6}. Therefore, the band topology change can be directly probed experimentally by the AHE. Similar to the case of GdZn, we have also investigated the topological properties and the intrinsic AHE in GdCd. As shown in Fig. \ref{fig4}(b), the value of the AHC for GdCd is $-$1300 $\Omega^{-1}$cm$^{-1}$ at 0.27 eV above the Fermi level and $-$653 $\Omega^{-1}$cm$^{-1}$ at the Fermi level. Our theoretical results on GdZn and GdCd need to be confirmed by future experimental studies.

\begin{figure}[htb]
\includegraphics[angle=0,scale=0.22]{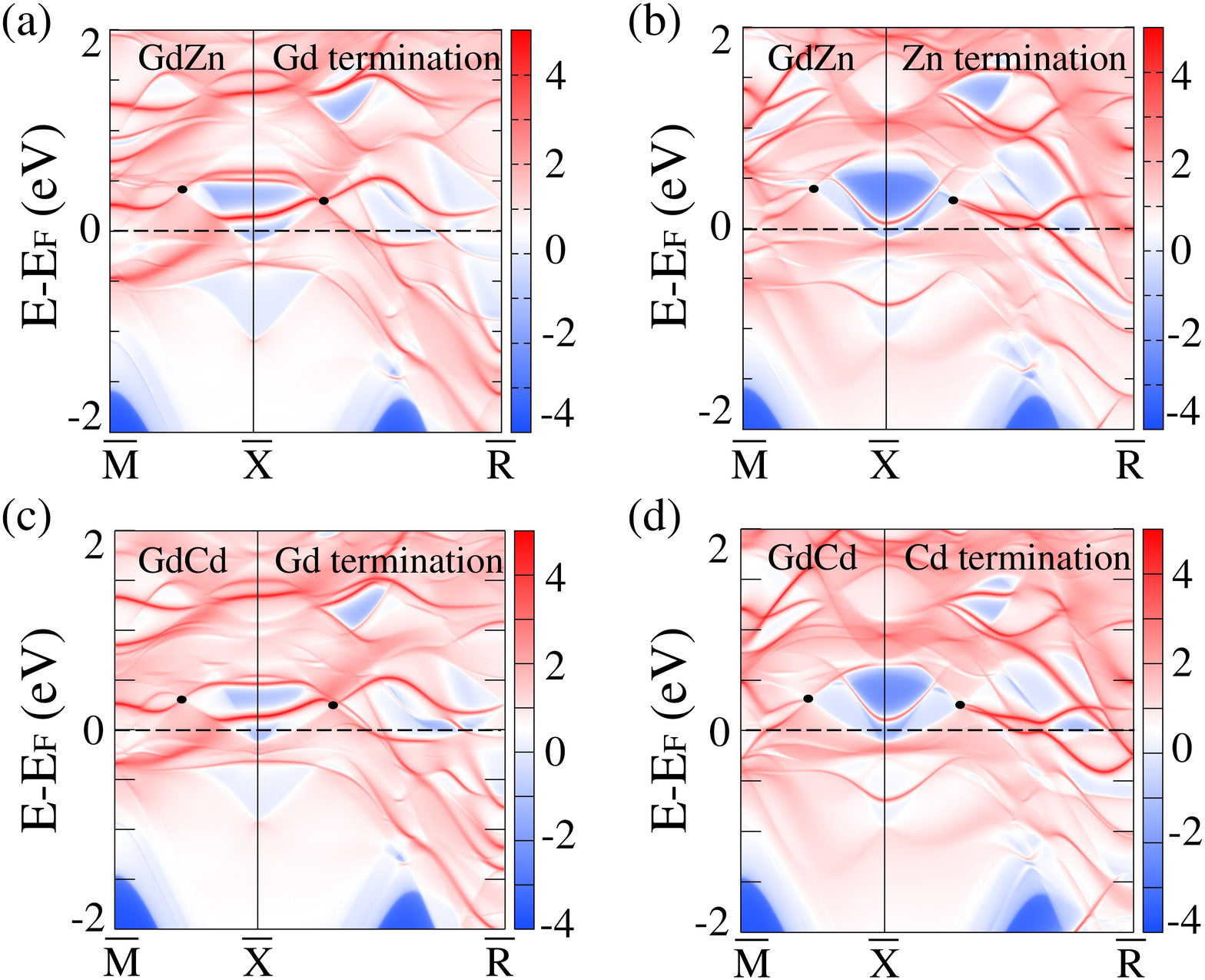}
  \caption{Projected band structures of the (101) surfaces along the high-symmetry lines [Fig.\ref{fig1}(b)] for the Gd-terminated and Zn/Cd-terminated (101) surfaces of (a)(b) GdZn and (c)(d) GdCd, respectively. The magnetization was set along the [001] direction (z-axis). }
 \label{fig5}
\end{figure}

The topological properties of GdZn and GdCd can be further verified by the nontrivial surface states. A significant feature of the topological nodal-line material is of the drumhead surface state\cite{Cu3PdN1,Cu3PdN2}. Figures \ref{fig5}(a)(b) and Fig. \ref{fig5}(c)(d) show the respective band structures of the (101) surfaces of GdZn and GdCd along the high-symmetry paths of the projected 2D BZ [Fig. \ref{fig1}(b)], which were calculated with the magnetization parallel to the [001] direction. The surface states around the \textit{$\bar{\mathrm{X}}$} point connect to the projected nodal points (black dots) within the $\bar{\mathrm{M}}-\bar{\mathrm{X}}$ and $\bar{\mathrm{X}}-\bar{\mathrm{R}}$ lines although part of them merge into the bulk states\cite{TlTaSe2,CBP}. This indicate that the nodal lines are topologically nontrivial.

\section{Discussion and Summary}

In previous studies, several kinds of ferromagnetic and ferrimagnetic materials with the Weyl points near the Fermi level were reported to show large intrinsic AHE\cite{Co3Sn2S2-1,Co3Sn2S2-2,Ti2MnAl,xugang,TbPtBi,PrAlGe-1,PrAlGe-2,Gd2C,Perovskites}. By contrast, the reported ferromagnetic nodal line materials with a large intrinsic AHE are rare, which include van der Waals semimetal Fe$_{3}$GeTe$_{2}$\cite{Fe3GeTe2} and Heusler compounds $A_2BC$ [Fe$_{2}$Mn$X$ ($X$ = P, As, Sb)\cite{Fe2MnP}, Co$_{2}$MnAl\cite{Co2MnAl}, Co$_2$VGa\cite{Co2VGa}, Co$_2$MnGa\cite{Co2MnGa-2}, Co$_2$TiSn\cite{Co2TiSn-1}]. Aa a new type of topological nodal line ferromagnets, GdZn and GdCd have many advantages. First, the Curie temperatures of GdZn (270 K) \cite{GdZn-TC} and GdCd (265 K) \cite{GdCd-TC} are both higher than that of Fe$_{3}$GeTe$_{2}$ (220 K)\cite{Fe3GeTe2}, which facilitates their experimental observation. Second, the intrinsic AHC of GdZn (GdCd) resulting from the gapped nodal lines can reach $-$ 1260 ($-$1300) $\Omega^{-1}$cm$^{-1}$ at 0.315 (0.270) eV above the Fermi level, where the Fermi level can be easily tuned into the nodal line gaps by alloying \cite{alloy}.  Third, since GdZn and GdCd are isotropic ferromagnets, the anisotropy of the AHC can be easily controlled by the magnetization direction (Fig. \ref{fig4}). Last but not least, the large Berry curvature may also induce the anomalous thermal Hall effect and anomalous Nernst effect\cite{Thermoelectric}.

In summary, based on the first-principles electronic structure calculations and the symmetry analysis, we have theoretically investigated the CsCl-type rare-earth intermetallics GdZn and GdCd. Due to the high crystal symmetry, these two materials host multiple nodal lines in the absence of SOC. The magnetization of GdZn and GdCd breaks the specified mirror symmetries, resulting in the gap opening of the corresponding nodal lines. The gapped nodal lines can create a huge Berry curvature and in turn generate a large intrinsic AHE. Our theoretical studies on GdZn and GdCd not only demonstrate the interplay between the crystal symmetry, the SOC effect, and the magnetism, but also provide a promising material platform for future transport and surface studies.


\begin{acknowledgments}

This work was supported by the National Key R\&D Program of China (Grants No. 2017YFA0302903 and No. 2019YFA0308603), the National Natural Science Foundation of China (Grants No. 11774422 and No. 11774424), the CAS Interdisciplinary Innovation Team, the Fundamental Research Funds for the Central Universities, and the Research Funds of Renmin University of China (Grant No. 19XNLG13).  N.Z. was supported by the Outstanding Innovative Talents Cultivation Funded Programs 2021 of Renmin University of China. Computational resources were provided by the Physical Laboratory of High Performance Computing at Renmin University of China.

\end{acknowledgments}

\end{document}